\documentclass[twocolumn,showkeys,prc,floatfix,nofootinbib]{revtex4}

\usepackage{amsmath}
\usepackage{amssymb}
\usepackage{graphicx}
\usepackage{epsfig}
\usepackage{natbib}

\def\mnras{Mon. Not. Roy. Astr. Soc.}
\def\apj{Astrophys. J.}

\def\aap{A. \& A.}
\def\plb{Phys. Lett. B}
\def\pr{Phys. Rev.}

\begin{document}

\title{The relativistic Feynman-Metropolis-Teller treatment at finite temperatures}

\author{S.~M.~de Carvalho}
\email{sheyse.martins@icra.it}
\affiliation{Dipartimento di Fisica and ICRA, Sapienza Universit\`a di Roma, P.le Aldo Moro 5, I--00185 Rome, Italy}
\affiliation{ICRANet, P.zza della Repubblica 10, I--65122 Pescara, Italy}

\author{M.~Rotondo}
\email{michael.rotondo@icra.it}
\affiliation{Dipartimento di Fisica and ICRA, Sapienza Universit\`a di Roma, P.le Aldo Moro 5, I--00185 Rome, Italy}
\affiliation{ICRANet, P.zza della Repubblica 10, I--65122 Pescara, Italy}

\author{Jorge A.~Rueda}
\email{jorge.rueda@icra.it}
\affiliation{Dipartimento di Fisica and ICRA, Sapienza Universit\`a di Roma, P.le Aldo Moro 5, I--00185 Rome, Italy}
\affiliation{ICRANet, P.zza della Repubblica 10, I--65122 Pescara, Italy}

\author{R.~Ruffini}
\email{ruffini@icra.it}
\affiliation{Dipartimento di Fisica and ICRA, Sapienza Universit\`a di Roma, P.le Aldo Moro 5, I--00185 Rome, Italy}
\affiliation{ICRANet, P.zza della Repubblica 10, I--65122 Pescara, Italy}

\date{\today}

\begin{abstract}

The Feynman-Metropolis-Teller treatment of compressed atoms has been recently generalized to relativistic regimes and applied to the description of static and rotating white dwarfs in general relativity. We present here the extension of this treatment to the case of finite temperatures and construct the corresponding equation of state (EOS) of the system; applicable in a wide regime of densities that includes both white dwarfs and neutron star outer crusts. We construct the mass-radius relation of white dwarfs at finite temperatures obeying this new EOS and apply it to the analysis of ultra low-mass white dwarfs with $M\lesssim 0.2 M_\odot$. In particular, we analyze the case of the white dwarf companion of PSR J1738+0333. The formulation is then extrapolated to compressed nuclear matter cores of stellar dimensions, systems with mass numbers $A\approx (m_{\rm Planck}/m_n)^3$ or mass $M_{\rm core}\approx M_{\odot}$, where $m_{\rm Planck}$ and $m_n$ are the Planck and the nucleon mass. For $T \ll m_e c^2/k_B \approx 5.9\times 10^9$ K, a family of equilibrium configurations can be obtained with analytic solutions of the ultra-relativistic Thomas-Fermi equation at finite temperatures. Such configurations fulfill global but not local charge neutrality and have strong electric fields on the core surface. We find that the maximum electric field at the core surface is enhanced at finite temperatures with respect to the degenerate case.

\end{abstract}


\maketitle
%
\section{Introduction}\label{sec:1}

We have recently generalized in Ref.~\cite{2011PhRvC..83d5805R} to relativistic regimes the classic work of Feynman, Metropolis and Teller (FMT)
\cite{feynman49}, solving a compressed atom by the Thomas-Fermi equation in a Wigner-Seitz cell.
The integration of this equation does not admit any regular solution for a point-like nucleus and both the nuclear radius 
and the nuclear composition have necessarily to be taken into account \cite{ferreirinho80,ruffini81}. This introduces a 
fundamental difference from the non-relativistic Thomas-Fermi model where a point-like nucleus is adopted.
So, this approach improves in the following aspects all previous treatments of the equation of state (EOS) of a compressed atom, including the classic works based on the uniform approximation by Chandrasekhar \cite{chandrasekhar31} and the EOS by Salpeter \cite{salpeter61}: 1) in order to guarantee 
self-consistency with a relativistic treatment of the electrons, the point-like assumption of the nucleus is 
abandoned introducing a finite sized nucleus; 2) the Coulomb interaction energy is fully
calculated without any approximation by solving numerically the relativistic Thomas-Fermi equation 
for each given nuclear composition; 3) the inhomogeneity of the electron distribution inside each Wigner-Seitz cell, 4) the energy-density of the system is calculated taking into account the contributions of the nuclei, of the Coulomb interactions, as well as of the relativistic electrons to the energy of the Wigner-Seitz cells; 5) the $\beta$-equilibrium between neutrons, protons and electrons is also taken into account leading to a self-consistent calculation of the threshold density for triggering the inverse $\beta$-decay of a given nucleus. The computation of the EOS is done by calculating the dependence of all these ingredients on the level of compression inside the star interior.

We have shown in Ref.~\cite{2011PhRvD..84h4007R} how all these effects together with general relativity are important in the determination of the macroscopic structure of white dwarfs as well as for the determination of their maximum stable mass against gravitational collapse. More recently, the relativistic FMT EOS has been used to determine general relativistic equilibrium configurations of rotating white dwarfs \cite{2013ApJ...762..117B}.

\begin{figure}[!hbtp]
\centering\includegraphics[width=\hsize,clip]{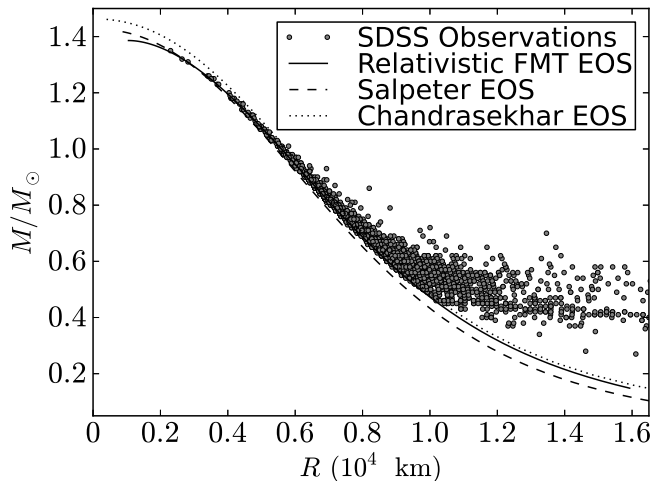}
\caption{Mass-radius relations of white dwarfs obtained with the relativistic FMT (solid black), Salpeter (dashed black), and Chandrasekhar (dotted black) EOS and their comparison with the estimated masses and radii of white dwarfs taken from the Sloan Digital Sky Survey Data Release 4 (SDSS-E06 catalog, gray circles) \cite{2011ApJ...730..128T}.}
\label{fig:SSDSwd}
\end{figure}

In Fig.~\ref{fig:SSDSwd} we show the mass-radius relation of $T=0$ white dwarfs for the relativistic FMT, Salpeter, and Chandrasekhar EOS and compare them with the estimated masses and radii of white dwarfs from the Sloan Digital Sky Survey Data Release 4 (SDSS-E06 catalog) \cite{2011ApJ...730..128T}. It can be clearly seen that already for masses $\lesssim 0.7$--$0.8~M_\odot$ deviations from the degenerate treatments are evident. It is natural to expect that such deviations could be related to the neglected effects of finite temperatures on the structure of the white dwarf. Thus, besides being interesting by their own, the finite temperature effects on the EOS and consequently on the mass-radius relation of the white dwarf are very important. In this work we extend our previous EOS \cite{2011PhRvC..83d5805R}, based on the degenerate relativistic FMT treatment, by introducing the effects of finite temperatures and use it to construct equilibrium configurations of white dwarfs at finite temperatures.

It is very interesting that there have been recently discovered ultra-low mass white dwarfs with masses $\lesssim 0.2 M_\odot$, which are companions of neutron stars in relativistic binaries; see e.g.~Refs.~\cite{2013Sci...340..448A,2012MNRAS.423.3316A}. These low-mass white dwarfs represent the perfect arena to testing the EOS of compressed matter since the central densities of these objects are expected to be $\lesssim 10^6$ g cm$^{-3}$, where the degenerate approximation breaks down and so temperature effects cannot be neglected. Using the mass-radius relation at finite temperatures, we analyze in the present work the structure of the white dwarf orbiting the pulsar PSR J1738+0333. We infer its mass, radius, surface gravity, internal temperature, and compare and contrast them with previous estimates.

The generalization of the relativistic FMT model presented in this work will be also useful to extend previous works in which the non-relativistic Thomas-Fermi model has been used to describe the physics of the low density layers of neutron stars including their atmospheres (see e.g.~Ref.~\cite{1998ApJ...502..847T}). The proper treatment of the relativistic and Coulomb effects corrects the over and underestimate of the total pressure at high and low densities respectively, which occurs in non-relativistic Thomas-Fermi models and in the approximate Coulomb corrections of Salpeter \cite{salpeter61}; see \cite{2011PhRvC..83d5805R}, for further details.

In addition to the generalization of the EOS of compressed matter, we follow the steps in \cite{2011PhRvC..83d5805R} and extrapolate the treatment to the case of compressed nuclear matter cores of stellar dimensions introduced in : macroscopic cores composed of neutrons, protons, and electron in $\beta$-equilibrium and with mass numbers $A\sim(m_{\rm Planck}/m_n)^3 \sim 10^{57}$, hence masses $M_{\rm core}\sim M_{\odot}$; expected to be bound by self-gravity. These objects are idealized configurations that reflect the properties of macroscopic nuclear matter systems such as neutron stars.

The paper is organized as follows: first in Sec.~\ref{sec:2} we describe the extension of the relativistic FMT treatment to finite temperatures. Then in Sec.~\ref{sec:3} we summarize the results of the numerical integration of the equations and describe the general properties of the new EOS. In Sec.~\ref{sec:4} we construct the mass-radius relation of white dwarfs and show specifically the results for $^4$He composition and in Sec.~\ref{sec:5} we apply these results to the case of the ultra-low mass white dwarf companion of PSR J1738+0333. In Sec.~\ref{sec:6} we extend the formulation of compressed matter to the case of the nuclear matter cores of stellar dimensions introduced in \cite{2011PhRvC..83d5805R}. We finally discuss our results in Sec.~\ref{sec:7}.

\section{The relativistic FMT treatment at finite temperatures}\label{sec:2}

We now consider the equations of equilibrium of a relativistic gas of electrons at a temperature $T\neq 0$ surrounding a finite sized and positively charged nucleus of mass and atomic numbers $A$ and $Z$, respectively. The electron cloud is confined within a radius $R_{\rm WS}$ of a globally neutral Wigner-Seitz cell and the system is isothermal.

Following Ref.~\cite{2011PhRvC..83d5805R}, we adopt a constant distribution of protons confined in a radius $R_c= \Delta \lambda_{\pi} Z^{\frac{1}{3}}$, where $ \lambda_{\pi}= \hbar / (m_\pi c)$ is the pion Compton wavelength, with $m_\pi$ the pion rest-mass. The parameter $\Delta$ is such that at nuclear density, $\Delta \approx (r_0 / \lambda_{\pi})(A/Z)^{1/3}$, where $r_{0} \approx 1.2$ fm; so in the case of ordinary nuclei $\Delta\approx 1$. Consequently, the proton number density can be written as 
\begin{equation}
 n_p (r)=\frac{3 Z}{4 \pi R_c^3} \theta(r-R_c)=\frac{3}{4 \pi \lambda^3_\pi\Delta^3} \theta(r-R_c),
\label{eq:eq1}
\end{equation}
where  
$\theta(r-R_{c})$ is the Heaviside function centered at the core (nucleus) radius, $r=R_c$.

Clearly, the electron number density follows from Fermi-Dirac statistics and is given by
\begin{equation}
n_e=\frac{2}{(2\pi \hbar)^{3}} \int_0^\infty \frac{4\pi p^{2} dp}
{\exp\left[\frac{\tilde{E}(p)- \tilde{\mu}_{e}(p)}{k_B T}\right] +1} ,
\label{eq:ne}
\end{equation}
where $k_B$ is the Boltzmann constant, $\tilde{\mu}_{e}$ is the electron chemical potential without the rest-mass, 
and $\tilde{E}(p)= \sqrt{c^2 p^2+m^2_e c^4} - m_{e}c^2$, with $p$ and $m_e$ the electron momentum and rest-mass, respectively.

Introducing the degeneracy parameter $\eta=\tilde{\mu}_e/(k_B T) $, $t=\tilde{E}(p)/(k_{B}T)$, and $\beta=k_{B}T/(m_e c^2)$, we can write the 
electron number density as
\begin{equation}
n_{e}=\frac{8\pi \sqrt{2}}{(2 \pi \hbar)^3} m^3 c^3 \beta^{3/2} \left[ F_{1/2} (\eta,\beta) + \beta F_{3/2} (\eta,\beta) \right],
\label{eq:ne2}
\end{equation}
where 
\begin{equation}
F_{k} (\eta,\beta)\equiv \int_{0} ^{\infty} \frac{t^k \sqrt{1+(\beta/2)t}}{1+ e^{t-\eta}}\, dt
\end{equation}
is the relativistic Fermi-Dirac integral.

We consider temperatures that satisfy $T\ll m_e c^2/k_B\approx 6\times 10^9$ K, so we will not take into account the presence of anti-particles. The Thomas-Fermi equilibrium condition for the relativistic electron gas is in this case given by
\begin{equation}\label{eq:TFcond2}
\tilde{\mu}_{e}(r) - e V(r) = k_B T \eta(r) -e V (r) ={\rm constant},
\end{equation}
where $V(r)$ is the Coulomb potential.

By introducing the dimensionless quantities $ x= r/\lambda_{\pi}$, $ x_c= R_c/\lambda_{\pi}$, $\chi/r = \tilde{\mu}_e/(\hbar c)$ and
replacing the above particle densities into the Poisson Equation
\begin{equation}
\nabla^2 V(r)= 4\pi e [n_p(r)-n_e(r)],
\label{eq:Poisson}
\end{equation}
we obtain the generalization of the relativistic Thomas-Fermi equation to finite temperatures
\begin{widetext}
\begin{equation}
 \frac{d^{2} \chi(x)}{dx^{2}}= -4\pi \alpha x \left\{ \frac{3}{4\pi \Delta^3} \theta(x_c-x) - \frac{\sqrt{2}}{\pi^2} \left(\frac{m_{e}}{m_{\pi}}\right) ^3 \beta^{3/2}\left[F_{1/2} (\eta,\beta) + \beta F_{3/2} (\eta,\beta)\right] \right\}.
\label{eq:eq4}
\end{equation}
\end{widetext}

The Eq.~(\ref{eq:eq4}) must be integrated subjected to the same boundary conditions as in the degenerate case, given by
\begin{equation}\label{eq:boundaryconds}
\chi(0)=0,\quad \left.\frac{d\chi}{dx}\right|_{x=0}>0, \quad \left.\frac{d\chi}{dx}\right|_{x=x_{\rm WS}}=\frac{\chi(x_{\rm WS})}{x_{\rm WS}},
\end{equation}
where the latter condition ensures the global charge neutrality at the Wigner-Seitz cell radius, $R_{\rm WS}$, and $x_{\rm WS}=R_{\rm WS}/ \lambda_\pi$ is the dimensionless cell radius. 

We turn now to compute the energy of the Wigner-Seitz cell. For the present case of finite temperatures, the total energy of each cell can be split as
\begin{equation}
E_{\rm WS}= E_{N}+E_k+E_{C} ,
\label{eq:etot}
\end{equation} 
where
\begin{align}
E_{N}&= M_{N}(A,Z)c^2 + U_{\rm th},\quad U_{\rm th}=\frac{3}{2} k_B T,\label{eq:EN}\\
E_{k}&= \int_{0} ^{R_{\rm WS}} 4\pi r^{2} ( {\cal E}_e - m_{e} n_{e}) dr,\label{eq:ek}\\
E_{C}&=\frac{1}{2} \int_{R_{c}} ^{R_{\rm WS}}  4\pi r^{2} e[n_{p}(r)-n_{e}(r)] V(r) dr,\label{eq:Ec}
\end{align}
are the nucleus, kinetic, and Coulomb energy. For the nucleus mass $M_{N}(A,Z)$ we adopt experimental values, $U_{\rm th}$ is the thermal energy of nuclei which we here adopt as an ideal gas \footnote{Quantum corrections to the ideal behavior of the ions considered here can be straightforwardly included following previous treatments such as \cite{1996A&A...314.1024S,1998PhRvE..58.4941C,2000PhRvE..62.8554P}}, and the electron energy density ${\cal E}_e$ is given by
\begin{align}\label{eq:Ek2}
{\cal E}_{e} &=  m_{e} c^2 n_{e} \nonumber \\
&+ \frac{\sqrt{2}}{\pi^2 \hbar^3} m_{e}^4 c^5 \beta^{5/2} \left[ F_{3/2} (\eta,\beta) + \beta F_{5/2} (\eta,\beta) \right].
\end{align}

The total density and pressure are then given by
\begin{align}
\rho &= \frac{E_{\rm WS}}{c^2 V_{\rm WS}},\label{eq:rho}\\
P &= P_N+P_{e}, \label{eq:P}
\end{align}
where
\begin{align}
P_N&= \frac{2}{3}\frac{U_{\rm th}}{V_{\rm WS}}=\frac{k_B T}{V_{\rm WS}} , \label{eq:PN} \\
P_e&= \frac{2^{3/2}}{3 \pi^2 \hbar^3} m_{e}^4 c^5 \beta^{5/2} \left[ F_{3/2} (\eta_{\rm WS},\beta)\right. \nonumber \\
&+ \left. \frac{\beta}{2} F_{5/2} (\eta_{\rm WS},\beta) \right],
\label{eq:Pe2}
\end{align}
being $\eta_{\rm WS}$ the value of $\eta$ at the boundary of the Wigner-Seitz cell with volume $V_{\rm WS}=4 \pi R_{\rm WS}^3/3$.

\section{Numerical integration of the equations and the EOS}\label{sec:3}

For a given chemical composition $(Z,A)$, temperature $T$ (i.e.~$\beta$), and dimensionless Wigner-Seitz cell radius $x_{\rm WS}$, the relativistic Thomas-Fermi equation (\ref{eq:eq4}) is integrated subjected to the boundary conditions (\ref{eq:boundaryconds}). We thus obtain both the Coulomb potential and the function $\eta$ inside the given Wigner-Seitz cell. With the knowledge of $\eta_{\rm WS}$, we proceed to evaluate first the energy of the cell by Eqs.~(\ref{eq:etot}--\ref{eq:Ek2}) and subsequently the values of the density and pressure through Eqs.~(\ref{eq:rho}--\ref{eq:Pe2}). For fixed chemical composition and temperature, we repeat the above steps for different cell radii to obtain different compression levels of the system; this leads to different densities and pressures, hence the EOS. These steps can be then performed for different compositions and temperatures; the results are discussed below.

\subsection{Properties of the EOS}

\begin{figure}[!hbtp]
\centering
\includegraphics[width=\hsize,clip]{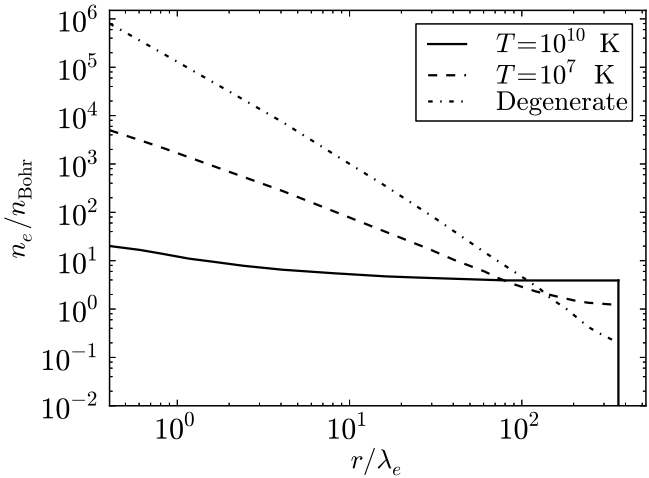}
\caption{Electron number density inside a Wigner-Seitz cell of $^{56}$Fe at a density of $30$ g cm$^{-3}$ at selected temperatures. Here $n_{\rm Bohr}=3/(4 \pi R^3_{\rm Bohr})\approx 1.6\times 10^{24}$ cm$^{-3}$, where $R_{\rm Bohr}=\hbar/(e^2 m_e)\approx 5.3\times 10^{-9}$ cm, is the Bohr radius. In this example we have used both low density and high temperatures up to $10^{10}$ K in order to show an extreme example of electron density flattening.}
\label{fig:neFe}
\end{figure}

As we showed in Ref.~\cite{2011PhRvC..83d5805R}, as a result of the Coulomb interaction duly accounted for in the relativistic Thomas-Fermi treatment, the distribution of the electrons inside a Wigner-Seitz cell is not uniform. In order to show the effects of the temperature, in Fig.~\ref{fig:neFe} we show as an example the electron number density inside a Wigner-Seitz cell of $^{56}$Fe at a density of $30$ g cm$^{-3}$ and for temperatures $T=[0,10^7,10^{10}]$ K.

\begin{figure}[!hbtp]
\centering
	\includegraphics[width=\hsize,clip]{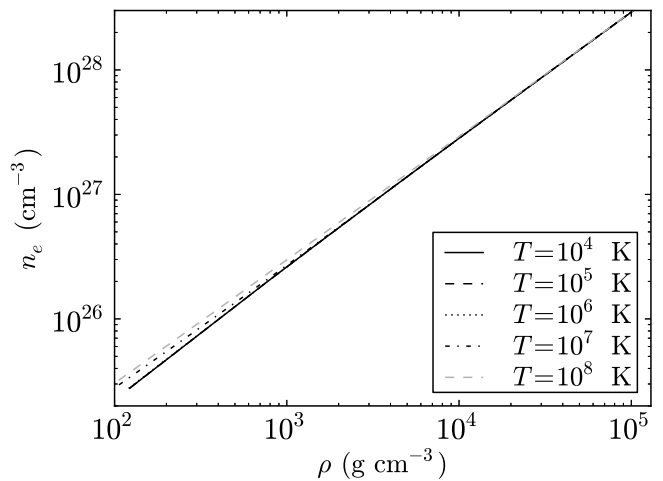}
\caption{Electron number density at the radius of a Wigner-Seitz cell of $^{12}$C as a function of the density (\ref{eq:rho}) for the selected temperatures $T=[10^4,10^5,10^6,10^7,10^8]$ K.}
\label{fig:rhone}
\end{figure}

We can see in Fig.~\ref{fig:neFe} how the effect of the temperature tends to homogenize the electron distribution inside the cell. In addition, we notice that the larger the temperature the larger the value of the electron density at the border of the Wigner-Seitz cell, thus increasing the electron pressure. This effect can be clearly seen in Fig.~\ref{fig:rhone} where we show the value of the electron number density evaluated at the cell radius, $R_{\rm WS}$, as a function of the density for the temperatures $T=[10^4,10^5,10^6,10^7,10^8]$ K, for a given chemical composition, $^{12}$C.

\begin{figure}[!hbtp]
\centering
\includegraphics[width=\hsize,clip]{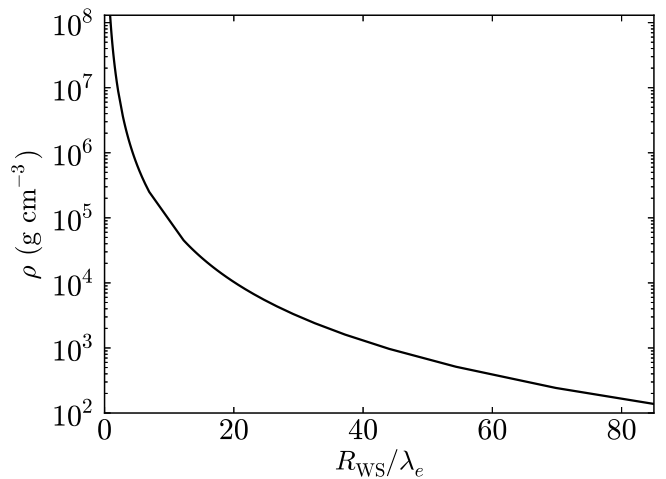}
\caption{Total density (in g cm$^{-3}$) of the system as a function of the radius of the Wigner-Seitz cell (in units of the electron Compton wavelength $\lambda_e=\hbar/(m_e c)\approx 3.9\times 10^{-11}$ cm) in the case of $^{12}$C at a temperature $T=10^7$ K.}
\label{fig:rhoRws}
\end{figure}

The volume of the Wigner-Seitz cell, $V_{\rm WS}=4 \pi R^3_{\rm WS}/3$, determines the density of the system $\rho$ given by Eq.~(\ref{eq:rho}); the smaller the volume the larger the density. In Fig.~\ref{fig:rhoRws} we show the density of the system as a function of the Wigner-Seitz cell radius $R_{\rm WS}$ for 
a temperature $T=10^7$ K and $^{12}$C chemical composition. Small deviations of the $R^{-3}_{\rm WS}$ behavior are due to the inhomogeneity of the electron distribution inside the cell and to the contribution of the Coulomb and electron kinetic energy to the density.

\begin{figure}[!hbtp]
\centering
	\includegraphics[width=\hsize,clip]{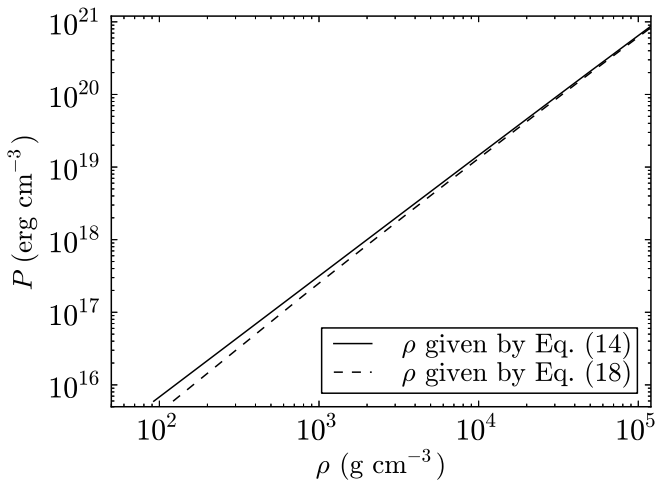}
\caption{Total pressure as a function of the matter density $\rho=A M_u n_e/Z$, given by Eq.~(\ref{eq:rhorest}), and $\rho=E_{\rm WS}/(c^2 V_{\rm WS})$ given by Eq.~(\ref{eq:rho}) which includes the thermal, kinetic, and Coulomb energy in the Wigner-Seitz cell. In this example the composition is $^{12}$C and the temperature $T=10^4$ K.}
\label{fig:diffrho}
\end{figure}

In this line it is important to mention that often in the literature the density of the system is approximated as 
\begin{equation}\label{eq:rhorest}
\rho=\frac{A}{Z} M_u n_e,
\end{equation}
which corresponds to the rest-mass density of nuclei in the system and where a uniform distribution of electrons is assumed. Here $M_u=1.6604\times 10^{-24}$ g is the unified atomic mass. We can see from Eq.~(\ref{eq:etot}) that this is equivalent to neglect the thermal, kinetic, and Coulomb energy of the cells as well as the inhomogeneity of the electron density. However, as we showed in Refs.~\cite{2011PhRvC..83d5805R,2011PhRvD..84h4007R}, the inclusion of the Coulomb and electron kinetic energies are important at low and high densities, respectively. In particular, the contribution of the kinetic energy of the electrons to the energy density is fundamental in the determination of the critical density for the gravitational collapse of $^{12}$C white dwarfs \cite{2011PhRvD..84h4007R}. We show in Fig.~\ref{fig:diffrho} the effect on the EOS of using as density of the system only the nuclei rest-mass, Eq.~(\ref{eq:rhorest}), instead of the full mass density given by Eq.~(\ref{eq:rho}) which accounts for the total energy of the Wigner-Seitz cell given by Eq.~(\ref{eq:etot}).

\begin{figure}[!hbtp]
\centering
	\includegraphics[width=\hsize,clip]{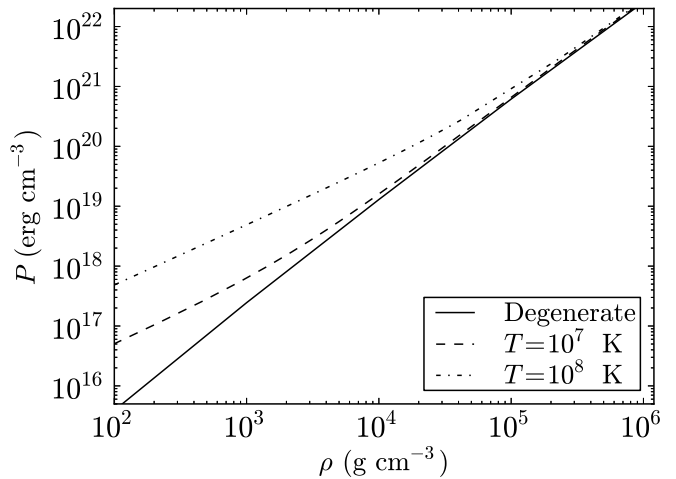}
\caption{Comparison of the EOS for $^{12}$C at temperatures $T=[0,10^7,10^8]$ K.}
\label{fig:EOScomparison}
\end{figure}

The effects of finite temperatures are clearly expected to be important at low densities, where the system looses its degeneracy. The point where the EOS should start to deviate from its degenerate behavior can be estimated by equating the degenerate and ideal gas pressures for the electron component. Assuming the electrons as non-relativistic we have, $n_e k_B T = (3 \pi^2)^{2/3}\hbar^2 n_e^{5/3}/m_e$, from which we obtain that temperature effects are important for densities
\begin{equation}\label{eq:Teffect}
\rho\lesssim 1.5\times 10^3 \left(\frac{T}{10^7\,{\rm K}}\right)^{3/2}\,{\rm g\,cm}^{-3}\, ,
%
\end{equation}
where we have used $A/Z\approx 2$ and $\rho\approx A M_u n_e/Z$. In Fig.~\ref{fig:EOScomparison} we compare the relativistic degenerate FMT EOS \cite{2011PhRvC..83d5805R,2011PhRvD..84h4007R} and its generalization at finite temperatures presented in this work, for the cases $T=10^7$ and $10^8$ K and $^{12}$C chemical composition. For these specific temperatures we see that deviations of the degenerate EOS start at a density $\rho\approx 2\times 10^4$ g cm$^{-3}$ and $\approx 10^6$ g cm$^{-3}$, respectively. For the same temperatures, Eq.~(\ref{eq:Teffect}) estimate deviations from degeneracy at $\rho\approx 1.5\times 10^3$ g cm$^{-3}$ and $\approx 4.8\times 10^4$ g cm$^{-3}$, respectively. Thus, the lower the temperature the better the estimate given by Eq.~(\ref{eq:Teffect}); the reason for this is that for larger temperatures the system will loose the degeneracy at larger densities where the non-relativistic approximation for the electrons breaks down.

\begin{figure}[!hbtp]
\centering
	\includegraphics[width=\hsize,clip]{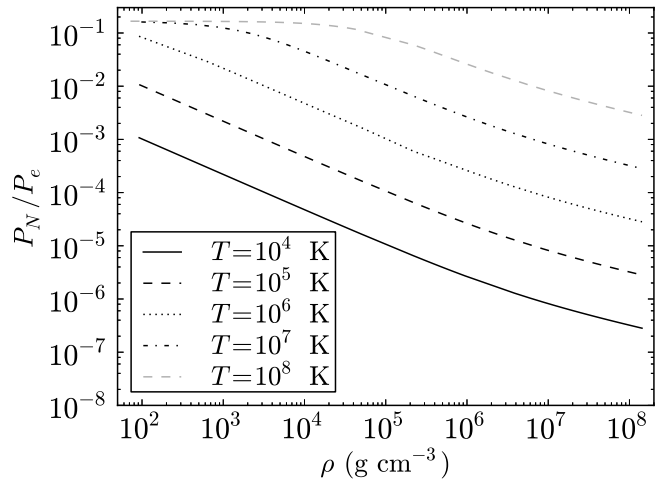}
\caption{Nuclei to electron pressure ratio as a function of the mass density in the case of $^{12}$ C white dwarf
for selected temperatures in the range $T=10^4$--$10^8$ K.}
\label{fig:PnPe}
\end{figure}

In Fig.~\ref{fig:PnPe}, we show the nuclei to electron pressure ratio in cells of $^{12}$C as a function of the density and for selected temperatures. It can be seen that for all temperatures the ratio approaches the same constant value in the low density regime. This is due to the fact that at sufficiently low densities also the electron gas becomes an ideal gas and consequently its pressure is approximately given by $P^{\rm id}_e =Z k_B T/V_{\rm WS}$. Therefore, the nuclei to pressure ratio approaches the limit $P_N/P^{\rm id}_e=1/Z$, where $P_N$ is given by Eq.~(\ref{eq:PN}). In the example of Fig.~\ref{fig:PnPe} we have $Z=6$ so $P_N/P^{\rm id}_e\approx 0.17$. It is clear that the density at which each curve reaches such a constant value increases with the temperature, since at larger temperatures the electrons reach their ideal gas state at higher densities.

\begin{figure}[!hbtp]
\centering
	\includegraphics[width=\hsize,clip]{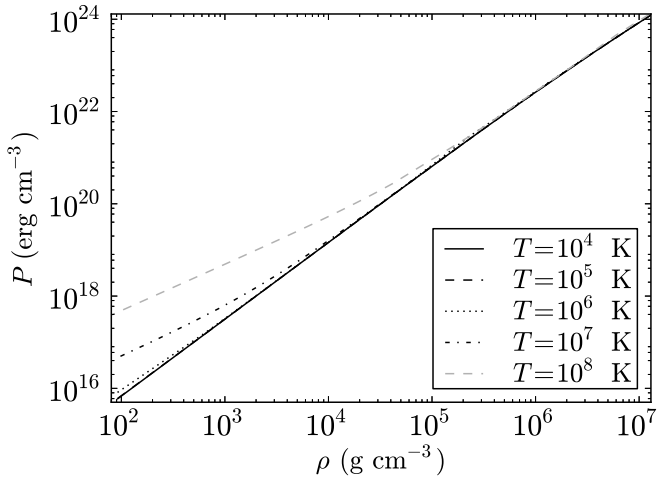}
\caption{Total pressure as a function of the mass density in the case of $^{12}$ C white dwarf
for selected temperatures in the range $T=10^4$--$10^8$ K.}
\label{fig:EOS}
\end{figure}

We summarize the finite temperature generalization of the relativistic FMT EOS in Fig.~\ref{fig:EOS}, where we plot as an example the total pressure (\ref{eq:P}) as a function of the total density of the system (\ref{eq:rho}) at temperatures $T=[10^4,10^5,10^6,10^7,10^8]$ K  and for a chemical composition, $^{12}$C. All the above features of the EOS are general and therefore applied also to chemical compositions other than $^{12}$C.

\subsection{Inverse $\beta$-decay and pycnonuclear reactions}

We turn now to the finite temperature effects on the inverse $\beta$-decay instability. It is known that white dwarfs may become unstable against the inverse $\beta$-decay process $(Z,A)\rightarrow (Z-1,A)$ through the capture of energetic electrons. In order to trigger such a process the electron energy must be larger than the mass difference between the initial nucleus $(Z,A)$ and the final nucleus $(Z-1,A)$. This threshold energy is denoted as $\epsilon^\beta _Z$. Usually it is satisfied $\epsilon^\beta _Z-1 < \epsilon^\beta _Z$ and therefore the initial nucleus undergoes two successive decays, i.e. $(Z,A)\rightarrow (Z-1,A) \rightarrow (Z-2,A)$; see e.g.~Refs.~\cite{1961ApJ...134..669S,1983JBAA...93R.276S}.

The critical density $\rho^{\beta} _{\rm crit}$ is then obtained numerically by looking for the density at which 
the electron energy equals $\epsilon^\beta _Z$. In Table II of Ref.~\cite{2011PhRvD..84h4007R} we showed that, in the degenerate case, the threshold energies to trigger the inverse $\beta$ process for $^{4}$He, $^{12}$C, $^{16}$O, and $^{56}$Fe are reached at densities, $\rho^{\beta} _{\rm crit}=1.37\times10^{11}$, $3.88\times 10^{10}$, $1.89\times 10^{10}$, and $1.14\times10^{9}$ g cm$^{-3}$, respectively. 

For the present finite temperature case, from our numerical integration we found that the critical densities for the occurrence of the inverse $\beta$-decay instability are not affected so that they are the same as in the degenerate approximation. This is due to the fact that the effects of temperatures $T\lesssim 10^8$ K become relevant at densities $\rho\lesssim 10^6$ g cm$^{-3}$, as can be seen from Figs.~\ref{fig:EOScomparison} and \ref{fig:EOS}.

We turn now to the pycnonuclear reactions. In a nuclei lattice the nuclear reactions proceed with the overcoming of the Coulomb barrier between neighbor nuclei. At zero temperatures, $T=0$, the Coulomb barrier can be overcome due to the zero-point energy of the nuclei (see e.g.~\cite{1969ApJ...155..183S,1983JBAA...93R.276S})
\begin{equation}\label{eq:omegap}
E_p=\hbar \omega_p\, ,\qquad \omega_p=\sqrt{\frac{4 \pi e^2 Z^2 \rho}{A^2 M_u^2}}\, .
\end{equation}

The number of pycnonuclear reactions per unit volume per unit time increases with the density of the system \cite{1969ApJ...155..183S} and any effect that reduces the Coulomb barrier will increase the cross-section of the reaction. The inclusion of the temperature could then lead to thermo-enhanced pycnonuclear rates (see e.g.~Refs.~\cite{1969ApJ...155..183S,2005PhRvC..72b5806G}). The astrophysical importance of pycnonuclear reactions e.g.~in the theory of white dwarfs relies on the fact that for instance the $^{12}$C+$^{12}$C pycnonuclear fusion, leading to $^{24}$Mg, is possible in a time scale shorter than a Hubble time, $\tau_{\rm pyc}<10$ Gyr, for densities $\sim 10^{10}$ g cm$^{-3}$. Such a density turns to be larger than the critical density $\sim 3\times 10^9$ g cm$^{-3}$ for the double inverse $\beta$-decay of $^{24}$Mg into $^{24}$Ne by electron capture (see e.g.~\cite{salpeter61,1983JBAA...93R.276S}), which destabilize the white dwarf due to sudden decrease of its electron pressure. Under such conditions, $^{12}$C+$^{12}$C fusion will indirectly induce the gravitational collapse of the white dwarf rather than to a supernova explosion.

Following the updated reaction rates of Ref.~\cite{2005PhRvC..72b5806G}, we recently computed in \cite{2013ApJ...762..117B} the critical density for pycnonuclear instability in general relativistic uniformly rotating $^{12}$C white dwarfs, at zero temperatures. It comes out that the instability agent of white dwarfs can be either general relativistic effects or inverse $\beta$-decay or pycnonuclear reactions or rotation through mass-shedding or secular instabilities (see \cite{2013ApJ...762..117B}, for details).

The electrons around the nuclei screen the positive charge of the nucleus reducing the Coulomb barrier; hence their proper inclusion could in principle increase the reaction rates. On the other hand, we showed in Figs.~\ref{fig:neFe} and \ref{fig:rhone} two different effects owing to the finite temperature: 1) it tends to flatten the electron distribution, thus changing the electron screening of the Coulomb potential with respect to the degenerate case; and 2) it increases the electron density hence the pressure at the border of the cell. These effects clearly could lead not only to qualitative but also to quantitative differences in the estimate of the rates of the pycnonuclear reactions (see e.g.~\cite{2012A&A...538A.115P}). 

However, the inclusion of these combined effects within the pycnonuclear reactions treatment, following a fully relativistic approach of the electron gas and the Coulomb interactions as the one presented here, is a most difficult and complex task that deserves a detailed and separated analysis, and therefore will not be addressed here.

\section{Mass-Radius relation}\label{sec:4}

General relativistic effects are important in the high density branch of white dwarfs; for instance they lead to the gravitational collapse of the star prior to the trigger of the inverse $\beta$-decay instability in $^{12}$C white dwarfs \cite{2011PhRvD..84h4007R}. We here construct the mass-radius relation of white dwarfs in their entire range of stability, so we use the equations of hydrostatic equilibrium within the framework of general relativity. Assuming the spherically symmetric metric
\begin{equation}\label{eq:metric}
ds^2 = e^{\nu(r)} c^2 dt^2 - e^{\lambda(r)}dr^2 - r^2 d\theta^2 - r^2 \sin^2 \theta d\varphi^2\, ,
\end{equation}
the equations of equilibrium can be written in the Tolman-Oppenheimer-Volkoff form
\begin{eqnarray}
\frac{d \nu(r)}{dr} &=& \frac{2 G}{c^2} \frac{4 \pi r^3 P(r)/c^2 + M(r)}{r^2 \left[1 - \frac{2 G M(r)}{c^2 r}\right]}\, ,\\
\frac{d M(r)}{dr} &=& 4 \pi r^2 \frac{{\cal E}(r)}{c^2}\, ,\\
\frac{d P(r)}{dr} &=& - \frac{1}{2} \frac{d \nu(r)}{dr} [{\cal E}(r)+P(r)]\, ,
\end{eqnarray}
where we have introduced the mass enclosed at the distance $r$ through $e^{-\lambda(r)} = 1 - 2 G M(r)/(c^2 r)$, ${\cal E}(r)=c^2 \rho(r)$ is the energy-density and $P(r)$ is the total pressure, given by Eqs.~(\ref{eq:rho}--\ref{eq:P}).

\begin{figure}[!hbtp]
\centering\includegraphics[width=\hsize,clip]{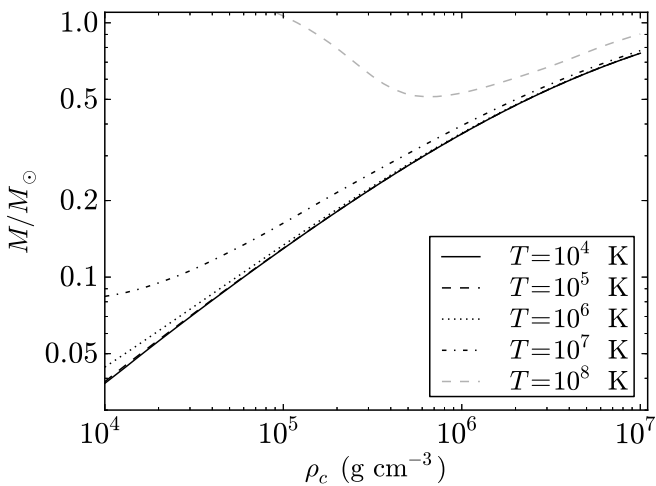}
\caption{Total mass versus central density for $^4$He white dwarfs for selected temperatures from $T=10^4$ K to $T=10^8$ K.}
\label{fig:mxrho}
\end{figure}

\begin{figure}[!hbtp]
\centering\includegraphics[width=\hsize,clip]{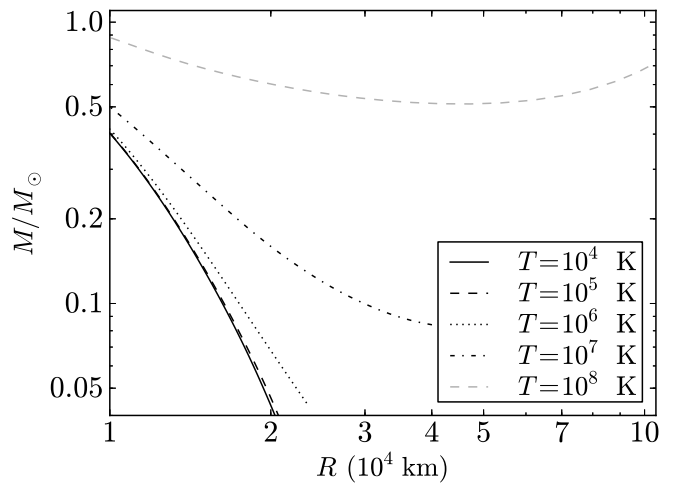}
\caption{Total mass versus radius for $^4$He white dwarfs for selected temperatures from $T=10^4$ K to $T=10^8$ K.}
\label{fig:mxr}
\end{figure}

These equations can be integrated for a wide range of central densities, temperatures, and for selected chemical compositions, for instance $^4$He, $^{12}$C, $^{16}$O, and $^{56}$Fe. In Figs.~\ref{fig:mxrho} and \ref{fig:mxr}, we show in particular the mass-central density and mass-radius relations of $^4$He white dwarfs in the range of densities and radii where finite temperature effects are more important.

The minima in these plots mark the transition from the ideal to the degenerate behavior of the electron gas: from left to right in the $M-\rho_c$ relation and from right to left in the $M-R$ relation. Thus these minima can be used to give an estimate of the minimum mass that a star should have to be able to burn stably a given chemical composition since the condition of a stable burning requires that the gas be non-degenerate. Consequently, stable burning requires that the star lies on the branch of solutions on the left-hand side of the minimum of the $M-\rho_c$ diagram or on the right-hand side of the minimum of the $M-R$ diagram. For instance, helium burning is triggered at a temperature $T_{\rm He+He}\approx 10^8$ K, so we can obtain from the solutions shown in Fig.~\ref{fig:mxrho} that the minimum mass for stable helium burning is $M^{\rm He+He}_{\rm min}\approx 0.51~M_\odot$. The corresponding radius and density of this configuration is $4.54\times 10^9$~cm$\approx 0.065~R_\odot$ and $6.59\times 10^5$~g~cm$^{-3}$, respectively. A similar analysis can be done for the other compositions.

\section{The ultra low-mass white dwarf companion of PSR J1738+0333}\label{sec:5}

It is clear that the effects of the temperature are particularly important at low densities, and hence for low-mass white dwarfs. We analyze here the specific case of the white dwarf companion of the millisecond pulsar PSR J1738+0333. We refer to \cite{2012MNRAS.423.3316A}, for details on the observations and technical aspects of the derivation of the binary parameters. 

Antoniadis et al.~\cite{2012MNRAS.423.3316A} obtained with the the Goodman High Throughput Spectrograph instrument of the Southern Astrophysical Research Telescope (SOAR) at Cerro Pach\'on, Chile, a photometric radius of the white dwarf, $R_{\rm WD}=0.042\pm0.004 R_\odot$. On the other hand, the analysis of the white dwarf atmosphere spectrum with the models of Ref.~\cite{2008arXiv0812.0482K} led to an effective surface temperature, $T_{\rm eff}=9130\pm 150$ K, and a logarithm of the surface gravity, $\log_{10}(g)=\log_{10}(G M_{\rm WD}/R^2_{\rm WD})= 6.55 \pm 0.1$.
Using the evolutionary mass-radius relation of Painei et al.~\cite{2000A&A...353..970P}, the mass of the white dwarf was estimated in Ref.~\cite{2012MNRAS.423.3316A} to be $M_{\rm WD} = 0.181^{+0.007} _{-0.005}$ M$_\odot$, and a corresponding radius $R_{\rm WD}=0.037^{+0.004}_{-0.003}~R_\odot$, in agreement with the photometric value.

A first attempt to obtain the mass of the white dwarf can be done directly from the observed data by combining the spectral and photometric analysis. Assuming the photometric radius as the star radius the mass of the white dwarf would be $M_{\rm WD}=g R^2_{\rm WD}/G\approx 0.23\,M_\odot$, using the central values of $R_{\rm WD}$ and $g$, which is roughly consistent with the mass derived from the mass-radius relation of Ref.~\cite{2000A&A...353..970P}.

\begin{figure}[!hbtp]
\centering\includegraphics[width=\hsize,clip]{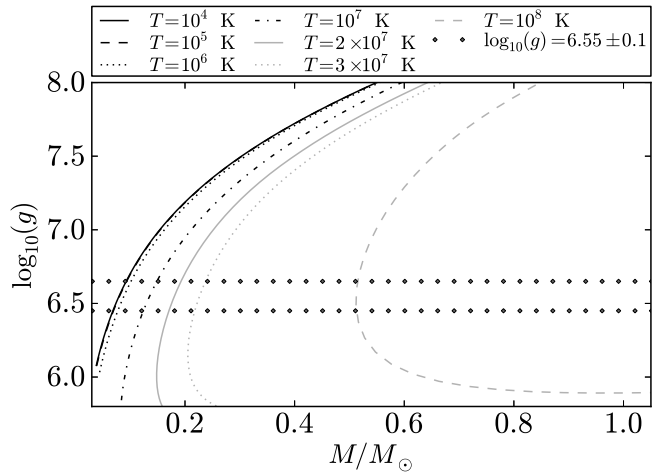}
\caption{Logarithm of the surface gravity, $\log_{10}(g)=\log_{10}(G M_{\rm WD}/R^2_{\rm WD})$, as a function of the mass for $^{4}$He white dwarfs for selected interior temperatures from $T=10^4$ K to $T=10^8$ K. The horizontal diamonds indicate the maximum and minimum best-fit values $\log_{10}(g)=6.55 \pm 0.1$.}
\label{fig:loggMHe}
\end{figure}

\begin{figure}[!hbtp]
\centering\includegraphics[width=\hsize,clip]{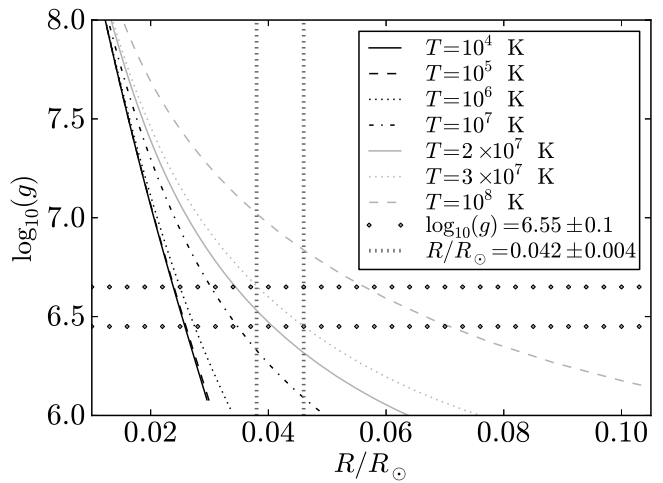}
\caption{Logarithm of the surface gravity, $\log_{10}(g)=\log_{10}(G M_{\rm WD}/R^2_{\rm WD})$, as a function of the radius for $^{4}$He white dwarfs for selected interior temperatures from $T=10^4$ K to $T=10^8$ K. The horizontal diamond markers and the vertical dashed markers indicate the maximum and minimum best-fit values of the surface gravity, $\log_{10}(g)=6.55 \pm 0.1$, and photometric radii $R_{\rm WD}= 0.042 \pm 0.004 R_\odot$, respectively.}
\label{fig:loggRHe}
\end{figure}

In order to compare our mass-radius relation at finite temperatures with the above results and infer the internal temperature of the white dwarf, we plotted in Figs.~\ref{fig:loggMHe} and \ref{fig:loggRHe} our theoretical surface gravity-mass and radius relations for $^4$He white dwarfs, together with the above observational constraints.

\begin{figure}[!hbtp]
\centering\includegraphics[width=\hsize,clip]{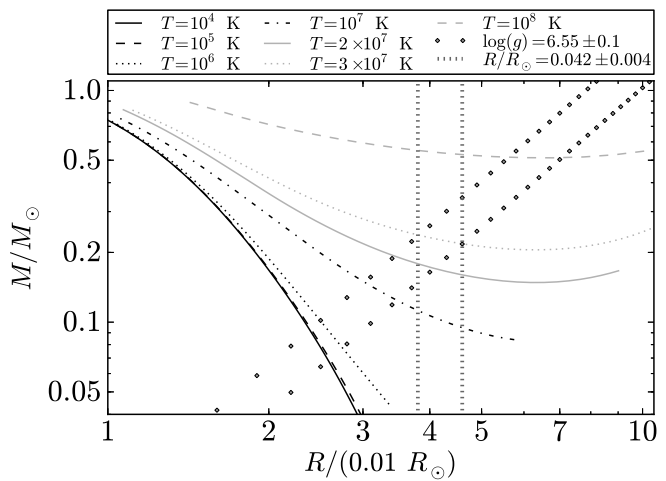}
\caption{Total mass versus radius for $^{4}$He white dwarfs for selected interior temperatures from $T=10^4$ K to $T=10^8$ K. The diagonal diamond markers and the vertical dashed markers indicate the maximum and minimum best-fit values of the surface gravity, $\log_{10}(g)=6.55 \pm 0.1$, and photometric radii $R_{\rm WD}= 0.042 \pm 0.004 R_\odot$, respectively.}
\label{fig:MRgHe}
\end{figure}

An inspection of Fig.~\ref{fig:loggMHe} does not give us any information on the possible internal temperature of the white dwarf since, in principle, we do not have any a priori information on the mass. However, from Fig.~\ref{fig:loggRHe} we clearly identify that the interior temperature of the white dwarf core should be $T \approx 2$--$3\times 10^7$ K. In Fig.~\ref{fig:MRgHe} we plot the mass-radius relation for $^{4}$He white dwarfs with the observational constraints of the companion of PSR J1738+0333. We can now compare our results with an estimate obtained for instance using the relation found by Koester in Ref.~\cite{1976A&A....52..415K} between the central and surface temperatures of the white dwarf, $T^4_{\rm eff}/g=2.05\times 10^{-10} T_c^{2.56}$. Using the value $T_{\rm eff}=9130$ K and $\log_{10}(g)=6.55$, this relation gives $T_c\approx 2.6\times 10^7$ K, in full agreement with our inference. In this estimate we have neglected the contribution of the thickness of the envelope to the total surface radius of the white dwarf. However, this approximation does not introduce a large error since the envelope would be in this case at most $\sim 10^{-2}~R_{\rm WD}$ thick.

\section{Application to nuclear matter cores of stellar dimensions}\label{sec:6}


In Ref.~\cite{2011PhRvC..83d5805R} we extended the relativistic FMT model to what we have called nuclear matter cores of stellar dimensions: macroscopic objects composed by neutrons, protons, and electrons in $\beta$-equilibrium, with mass numbers $A\sim (m_{\rm Planck}/m_n)^3 \sim 10^{57}$, and corresponding masses $M_{\rm core} \sim M_{\odot}$. These systems are expected to represent idealized cores of macroscopic systems of nuclear matter kept bound by self-gravity, such as the cores of neutron stars. We now follow our treatment in Ref.~\cite{2011PhRvC..83d5805R}, and use the existence of scaling laws and proceed to the ultrarelativistic limit of the relativistic Thomas-Fermi equation at finite temperature given by Eq.~(\ref{eq:eq4}).


The $\beta$-equilibrium of $N_n=A-Z$ neutrons, $Z$ protons, and $Z$ electrons gives for massive cores $N_n\gg Z$. Typically, in these systems we have $A/Z\approx10^2$, so at nuclear density the neutron gas will have a Fermi energy $E_n^F$ of the order of
\begin{equation}\label{eq:TFn}
E^F_n\simeq \frac{(P^F_n)^2}{2 m_n}\simeq (3\pi^2)^{2/3}\frac{\hbar^2}{2 m_n}\left(A\frac{\rho_{\rm nuc}}{m_n}\right)^{2/3}\sim 60\,{\rm MeV},
\end{equation}
where we have used a nuclear density value $\rho_{\rm nuc}\approx 2.7\times 10^{14}$ g cm$^{-3}$, and $1-Z/A\approx 1$. Assuming a temperature such that $T\ll T^F_n = E_n^F/k_B\approx 7\times 10^{11}$ K, the neutron chemical potential $\mu_n$ can be expanded as
\begin{equation}
\mu_n=E_n^F\left[1-\frac{\pi^2}{12}\left(\frac{k_B T}{E_n^F}\right)^2-\frac{\pi^4}{80}\left(\frac{k_B T}{E_n^F}\right)^4+...\right].
\label{eq:UR1a}
\end{equation}

Correspondingly, the protons have Fermi energy $E_p^F \sim (Z/A)^{2/3}E^F_n \sim$ MeV, so for temperatures $k_B T\ll E_p^F\approx 1$ MeV, Eq.~(\ref{eq:UR1a}) applies also for protons
\begin{equation}
\mu_p=E_p^F\left[1-\frac{\pi^2}{12}\left(\frac{k_B T}{E_p^F}\right)^2-\frac{\pi^4}{80}\left(\frac{k_{B}T}{E_p^F}\right)^4+ ...\right].
\label{eq:UR1b}
\end{equation}

As a result, for temperatures $k_B T\lesssim 1$ MeV, both neutrons and protons can be treated as degenerate particles whereas in this limit electrons are semi-degenerate and ultrarelativistic. In the case of ordinary nuclei, due their high isospin symmetry ($A/Z\approx 2$), both neutrons and protons can be treated as degenerate particles until $T\approx (Z/A)^{2/3}E^F_n/k_B \sim 38$ MeV.

Since in the ultrarelativistic limit for electrons their kinetic energy $\epsilon$ is simply $pc$, the condition $\mu_e/(k_BT)\gg 1$ holds. Consequently the integral
\begin{equation}
I=\int_{0}^{\infty}{\frac{f(\epsilon) d\epsilon}{\exp{\left(\frac{\epsilon-\mu_e}{k_{B}T}\right)}+1}},
\label{eq:UR1}
\end{equation} 
with $f(\epsilon)=\epsilon^2$ appearing in the electron density given by Eq.~(\ref{eq:ne}) can be expanded as
\begin{widetext}
\begin{equation}
I=\int_{0}^{\mu_e}{f(\epsilon) d\epsilon}+2(k_{B}T)^2f'(\mu_e)\int_{0}^{\infty}{\frac{z}{e^z+1}dz}+\frac{1}{3}(k_{B}T)^4f'''(\mu_e)\int_{0}^{\infty}{\frac{z^3}{e^z+1}dz}+...,
\label{eq:UR2}
\end{equation}
\end{widetext}
where
\begin{equation}
\int_{0}^{\infty}{\frac{z^{x-1}}{e^z+1}dz}=(1-2^{1-x})\Gamma(x)\sum_{n=1}^{\infty}{\frac{1}{n^x}},
\label{eq:UR3}
\end{equation}
with $\Gamma$ the Gamma function and $\mu_e$ the chemical potential of electrons and a prime denotes derivative with respect to $\epsilon$.
We thus obtain the result
\begin{align}
I&=\int_{0}^{\mu_e}{f(\epsilon) d\epsilon}+\frac{\pi^2}{6}(k_{B}T)^2f'(\mu_e) + \nonumber \\ 
&+\frac{7\pi^4}{360}(k_{B}T)^4f'''(\mu_e)+...,
\label{eq:UR4}
\end{align}
and retaining only the first term in $T$ we have
\begin{equation}
I \approx \frac{\mu_e^3}{3}+\frac{\pi^2}{6}(k_B T)^2\mu_e.
\label{eq:UR5}
\end{equation}
As discussed in \cite{2011PhRvC..83d5805R}, for a nuclear massive core of stellar dimensions we can assume the plane-parallel 
approximation, which leads to the Poisson equation in the case of finite temperatures
\begin{equation}
\frac{d^2\hat \phi}{d\xi^2}=-\theta(\xi-\xi_c)+\hat\phi^3+s\hat\phi,
\label{eq:UR6}
\end{equation}
where $\phi=4^{1/3}(9\pi)^{-1/3} \chi \Delta/x$, $\hat x=k x$ where $k=\left(12/\pi\right)^{1/6}\sqrt{\alpha}\Delta^{-1}$, $\xi=\hat x- \hat x_c$, and $s=(2\pi^4)^{1/3}\Delta^2(k_B T)^2/(3^{4/3}m_{\pi}^2c^2)$. Notice that the above equation is the ultrarelativistic version of Eq.~(\ref{eq:eq4}) for semi-degenerate electrons and how in the limit $T\to 0$ ($s\to 0$), it leads to the ultrarelativistic Thomas-Fermi equation for fully degenerate massive cores obtained in \cite{2011PhRvC..83d5805R}.

The Coulomb potential is given by
\begin{equation}
eV(\xi)=\left(\frac{9\pi}{4}\right)^{1/3}\frac{1}{\Delta} m_\pi c^2 \hat \phi(\xi)-C,
\label{eq:v01T}
\end{equation}
with $C=(9\pi/4)^{1/3}\Delta^{-1} m_{\pi} c^2 \hat\phi(\xi_{\rm WS})$, the electric field is 
\begin{equation}
E(\xi)=-\left(\frac{3^5\pi}{4}\right)^{1/6}\frac{\sqrt{\alpha}}{\Delta^2}\frac{m_\pi^2 c^3}{e\hbar }  \frac{d\hat \phi}{d\xi},
\label{eq:v01e}
\end{equation}
and the electron number density is
\begin{align}
n_e(\xi) &=\frac {(m_{\pi}c^2)^3}{3\pi^2\hbar^3c^3}\left[
\left(\frac{9\pi}{4}\right)\frac{1}{\Delta^3}  \hat \phi^3(\xi)+ \right. \nonumber \\
&+\left. \frac{\pi^2}{2}\left(\frac{9\pi}{4}\right)^{1/3}\frac{1}{\Delta}\left(\frac{k_{B}T}{m_{\pi}c^2}\right)^2\hat \phi(\xi)\right].
\label{eq:elnd1T}
\end{align}

The global charge neutrality of the system imposes the boundary condition that the electric field vanishes at $\xi=\xi_{\rm WS}$. This implies $d\hat \phi/d\xi|_{\xi=\xi_{\rm WS}}=0$. The function $\hat \phi$ and its first derivative $d\hat \phi/d\xi$ must be continuous at the surface $\xi=0$ of the nuclear density core. 
This boundary-value problem  can be solved analytically and indeed Eq.~(\ref{eq:UR6}) has the first integral,
\begin{align}\label{eq:Phiprima}
&2\left(\frac{d\hat \phi}{d\xi}\right)^2 = \nonumber\\
&\left\{\begin{array}{ll} \hat \phi^4(\xi)+2s\hat \phi^2-4\hat \phi(\xi) +3-2s,&\quad \xi\leq 0, \\
\hat \phi^4(\xi)+2s\hat \phi^2-\hat \phi^4(\xi_{\rm WS})-2s\hat\phi^2(\xi_{\rm WS}),&\quad \xi>0,
\end{array}\right.
\end{align} 
with boundary conditions at $\xi=0$: 
\begin{align}
\hat \phi(0)&=\frac{\hat \phi^4(\xi_{\rm WS})+3}{4}+\frac{s}{2}\left[\hat \phi^2(\xi_{\rm WS})-1\right],\\
\left.\frac{d\hat \phi}{d\xi}\right|_{\xi=0} &=-\bigg\{\frac{\hat \phi^4(0)-\hat \phi^4(\xi_{\rm WS})}{2} + \nonumber\\
&+  s[\hat \phi^2(0)-\hat \phi^2(\xi_{\rm WS})]\bigg\}^{1/2}.
\label{eq:UR8}  
\end{align}

The solution of Eq.~(\ref{eq:Phiprima}) in the interior region $\xi\leq 0$ is then
\begin{equation}\label{eq:phiint}
\hat{\phi}(\xi)=1-(s+3)\left[1+\left(\frac{s+1}{2}\right)^{1/2}\sinh(\beta-\sqrt{s+3}\xi)\right]^{-1}\, ,
\end{equation}
with
\begin{equation}\label{eq:phiintcons}
\sinh\beta = \sqrt{\frac{2}{s+1}}\left\{\frac{11+\phi^4(\xi_{\rm WS})+2 s [\phi^2(\xi_{\rm WS})+1]}{1-\phi^4(\xi_{\rm WS})-2s[\phi^2(\xi_{\rm WS})-1]} \right\}.
\end{equation}

In the exterior region $\xi> 0$ the solution of Eq.~(\ref{eq:Phiprima}) is 
\begin{equation}\label{eq:phiext}
 \hat{\phi}(\xi)=\frac{\sqrt{-s+\sqrt{s^2+G}}}{\cos \left(
    {\rm am}\left[(s^2+G)^{1/4}(\xi-\xi_{\rm WS}),\frac{1}{2}+\frac{s}{2\hat{\phi}^2(\xi_{\rm WS})}\right]\right)},
\end{equation}
where $G=\hat{\phi}^4(\xi_{\rm WS})+2 s\hat{\phi}^2(\xi_{\rm WS})$. It can be seen again how in the limit $T\to 0$ ($s\to 0$), the solution at finite temperatures given by Eqs.~(\ref{eq:phiint}), (\ref{eq:phiintcons}), and (\ref{eq:phiext}) becomes its degenerate counterpart obtained in \cite{2011PhRvC..83d5805R}.

From Eqs.~(\ref{eq:UR8}) follows that the peak of the electric field at the surface of the core is larger than the corresponding 
value obtained for $T=0$. In fact we have, for any temperature $T> 0$ and level of compression $\xi_{\rm WS}\neq 0$
\begin{equation}
\left|\left(\frac{d\hat \phi}{d\xi}\right)_{\xi=0}\right|_{T>0}>\left|\left(\frac{d\hat \phi}{d\xi}\right)_{\xi=0}\right|_{T=0}.
\label{eq:UR9}  
\end{equation}

As in the degenerate case, in the limit $\xi_{\rm WS}\rightarrow 0$, the global charge neutrality $N_e=Z$ and 
the local charge neutrality $n_e=n_p$ are recovered and at the surface of the massive core no electrodynamical 
structure is present. 

The above analytic equations can be used only in the ultra-relativistic regime of the electron gas; it can then be checked from the above formulation that at such high compressions
we have $\hat \phi(\xi)|_{T>0}\approx \hat \phi(\xi)|_{T=0} $. More specifically, corrections due to thermal 
effects on the density of ultra-relativistic electrons are smaller than $1\%$ for $T\lesssim 0.1$ MeV$/k_B\approx 10^9$ K.

\section{Conclusions}\label{sec:7}

The Feynman-Metropolis-Teller treatment \cite{2011PhRvC..83d5805R} of compressed matter has been here generalized to the case of finite temperatures. We have thus obtained the EOS formed by nuclei and electrons by solving the finite temperature relativistic Thomas-Fermi equation (\ref{eq:eq4}) within globally neutral Wigner-Seitz cells. We emphasize in this work on the electron component and the Coulomb interaction between ions and electrons fully computed within a relativistic Thomas-Fermi approach with finite sized nuclei, and therefore applicable to any relativistic regime of the electrons and densities. This work generalizes other treatments based on either a uniform distribution of electrons or the classic Thomas-Fermi treatment; see e.g.~\cite{1998ApJ...502..847T}. The quantum corrections to the classic ideal ion fluid considered in this work can be straightforwardly introduced in their corresponding ranges of relevance as done in previous treatments; see e.g.~\cite{1996A&A...314.1024S,1998PhRvE..58.4941C,2000PhRvE..62.8554P,2013A&A...550A..43P}.

We have shown the general features of the new EOS and compared and contrasted the effects owing to the non-zero temperature with respect to the degenerate case. We have checked that the onset of the inverse $\beta$-decay instability is not modified for temperatures $T\lesssim 10^8$ K and therefore the zero-temperature critical densities computed in Ref.~\cite{2011PhRvD..84h4007R} can be safely used. The enhancement and flattening of the electron density inside the cell for larger temperatures could have relevant effect in the pycnonuclear reaction rates in the interior of white dwarfs and/or in the low density layers of accreting neutron stars. 

Deviations from the degenerate EOS have been shown to occur in the regions of interest of low-mass white dwarfs and in the outermost layers of neutron star crusts. Ultra-low mass white dwarfs, $M_{\rm WD}\sim 0.2 M_\odot$ \cite{antoniadisscience2013,2012MNRAS.423.3316A}, have been found in binary systems with neutron stars companions. These objects have central densities $\lesssim 10^6$ g cm$^{-3}$, where the degenerate approximation breaks down and so thermal effects cannot be neglected. We have analyzed here the specific case of PSR J1738+0333, whose mass and radius was estimated in \cite{2012MNRAS.423.3316A} using the evolutionary mass-radius relation of Painei et al.~\cite{2000A&A...353..970P}. They obtained $M_{\rm WD} = 0.181^{+0.007} _{-0.005}$ M$_\odot$, $R_{\rm WD}=0.037^{+0.004}_{-0.003}~R_\odot$, in agreement with the spectrometric and photometric data. We inferred for this object an internal temperature $T\approx 2$--$3\times 10^7$~K, and a mass $M_{\rm WD}\approx 0.2~M_\odot$ assuming for instance the photometric radius, $R=0.042~R_\odot$, as the star radius. We checked also our result using the relation by Koester \cite{1976A&A....52..415K} between the internal and surface white dwarf temperatures, $T^4_{\rm eff}/g=2.05\times 10^{-10} T_c^{2.56}$. Using the surface temperature and the logarithm of the surface gravity obtained from the spectral analysis, $T_{\rm eff}=9130$ K and $\log_{10}(g)=6.55$, this relation gives $T_c\approx 2.6\times 10^7$ K, in full agreement with our results.

Following our previous work \cite{2011PhRvC..83d5805R}, we finally extrapolated the treatment to macroscopic systems with mass numbers $A\approx (m_{\rm Planck}/m_n)^3\sim 10^{57}$, corresponding to masses $M_{\rm core}\approx M_{\odot}$. We showed that the presence of the temperature enhances the maximum electric field in the core surface of these objects.

\begin{acknowledgments}
S.~M. de Carvalho acknowledges the support given by the International Relativistic Astrophysics Erasmus Mundus Joint Doctorate Program under the Grant 2010--1816 from EACEA of the European Commission. We are grateful to the referee for the comments and suggestions which helped us to improve the presentation of our results.
\end{acknowledgments}




\end{document}